\documentclass[%
 reprint,
 amsmath,amssymb,
 aps,
]{revtex4-2}

\usepackage{graphicx}
\usepackage{dcolumn}
\usepackage{bm}
\usepackage{hyperref}

\begin{document}

\preprint{APS/123-QED}

\title{Tunable and Enhanced Rashba Spin-Orbit Coupling in Iridate-Manganite Heterostructures}

\author{T. S. Suraj$^{1,2}$} 
\thanks{Authors contributed equally}
\author{Ganesh Ji Omar$^{3,4}$}
\thanks{Authors contributed equally}
\author { Hariom Jani$^{4,6}$, Muhammad Mangattuchali Juvaid$^{1,3,4}$, Sonu Hooda$^4$, Anindita Chaudhuri$^5$, Andrivo Rusydi$^{3,5}$}
\author{Kanikrishnan Sethupathi$^2$}
\author{Thirumalai Venkatesan$^{3,4,6,7,8}$}
\author{Ariando Ariando$^{3,4,6}$}
\email{ariando@nus.edu.sg}
\author{Mamidanna Sri Ramachandra Rao$^1$}
\email{msrrao@iitm.ac.in}

 \affiliation{$^1$ Department of Physics, Nano Functional Materials Technology Center, Material Science Research Center, IIT Madras, India 600036}
 \affiliation{$^2$ Low Temperature Physics Lab, Department of Physics,Indian institute of Madras, Chennai, India 600036}%
 \affiliation{$^3$ Department of Physics, National University of Singapore, Singapore 117542, Singapore}
\affiliation{$^4$ NUSNNI-NanoCore, National University of Singapore, Singapore 117411}
\affiliation{$^5$ Singapore Synchrotron Light Source, National University of Singapore, 5 Research Link, Singapore 117603, Singapore}
\affiliation{$^6$ National University of Singapore Graduate School for Integrative Sciences and Engineering (NGS), University Hall, 21 Lower Kent Ridge Road, Singapore 119077}
\affiliation{$^7$ Department of Materials Science and Engineering, National University of Singapore, Singapore 117575, Singapore}
\affiliation{$^8$ Department of Electrical and Computer Engineering, National University of Singapore, Singapore 117576, Singapore}
\date{\today}

\begin{abstract}
Tailoring spin-orbit interactions and Coulomb repulsion are the key features to observe exotic physical phenomena such as magnetic anisotropy and topological spin texture at oxide interfaces. Our study proposes a novel platform for engineering the magnetism and spin-orbit coupling at LaMnO$_3$/SrIrO$_3$ (3\textit{d}-5\textit{d} oxide) interfaces by tuning the LaMnO$_3$ growth conditions which controls the lattice displacement and spin-correlated interfacial coupling through charge transfer. We report on a tunable and enhanced interface-induced Rashba spin-orbit coupling and Elliot-Yafet spin relaxation mechanism in LaMnO$_3$/SrIrO$_3$ bilayer with change in the underlying magnetic order of LaMnO$_3$. We also observed enhanced spin-orbit coupling strength in LaMnO$_3$/SrIrO$_3$ compared to previously reported SrIrO$_3$ layers. The X-Ray spectroscopy measurement reveals the quantitative valence of Mn and their impact on charge transfer. Further, we performed angle-dependent magnetoresistance measurements, which show signatures of magnetic proximity effect in SrIrO$_3$ while reflecting the magnetic order of LaMnO$_3$. Our work thus demonstrates a new route to engineer the interface induced Rashba spin-orbit coupling and magnetic proximity effect in 3\textit{d}-5\textit{d} oxide interfaces which makes SrIrO$_3$ an ideal candidate for spintronics applications.
\end{abstract}

\maketitle

\section{\label{sec:level1}Introduction}
The combination of artificially layered complex oxides in heterostructures opens the possibility of realizing novel functional properties from the strong interplay among charge, spin, orbit and lattice degrees of freedom which might be absent in the constituent oxide layers. Moreover, the interfacial effects mediated through charge transfer between oxide layers play a significant role in tuning the interface physics and its resultant properties \cite{1,2,3,4,5}. Among oxides, there is a surge in research interest for combinations of 3\textit{d}-5\textit{d} oxide interfaces for exploring various novel phenomena, such as manipulation of spin-orbit coupling that has potential applications in spintronics memory devices \cite{6,7,8}. Among 5{\textit{d}} oxide materials, iridates are the most exciting due to the combination of large intrinsic spin-orbit coupling (interaction strength; $\xi$)) and their tunable coulombic correlations (interaction strength; $ U $) \cite{9}. The strong spin-orbit coupling in 5{\textit{d}} orbital state splits the $t_{2g}$ levels due to crystal field into a fully filled $J_{eff}{= \frac {3}{2}}$ quartet and the $J_{eff}{=\frac {1}{2}}$ doublet having a single electron (hole) forming a half-filled band. Depending on the interaction strength ($U$), an iridate system can become a Mott insulator, or it can be driven to have a metallic/semi-metallic ground state \cite{10,11,12}. Perovskite SrIrO$_3$ can be epitaxially grown over various transition metal oxides (TMO), and its semi-metallicity can be tuned by compressive strain and reduced dimensionality, which makes it an ideal choice as 5{\textit{d}} oxide layers in 3\textit{d}-5\textit{d} heterostructures \cite{13,14,15}.

On the other hand, LaMnO$_3$ is the parent compound for manganite, containing the 3$d$ element Mn, which is an A-type antiferromagnetic insulator in bulk and could behave like a ferromagnet in epitaxial thin films due to vacancies or epitaxial strain \cite{16,17,18}. Our earlier studies demonstrate the origin of ferromagnetism in LaMnO$_3$/SrTiO$_3$ heterostructures by mapping the magnetic domains which show long-range ferromagnetic ordering arising from electron doping at the LaMnO$_3$/SrTiO$_3$ due to polar catastrophe \cite{19}. LaMnO$_3$ thin films grown under different deposition oxygen partial pressures ($pO_2$) have also been systematically studied by different groups with a variety of experimental techniques such as X-ray Absorption Spectroscopy (XAS), X-ray Magnetic Circular Dichroism (XMCD) and Transmission Electron Microscopy-Electron Energy Loss Spectroscopy (TEM-EELS) \cite{18,19,20,21,22,23}. Roqueta \textit{et al.} reported tunability of strain-controlled ferromagnetism in LaMnO$_3$ during growth by varying the background $pO_2$ that resulted in a rich magnetic phase diagram. However, the oxygen non-stoichiometry creates an imbalance in Mn valence states by charge ordering, which induces double exchange mediated ferromagnetism in LaMnO$_3$ \cite{20}. This aspect of oxygen non-stoichiometry tuning to control spin-orbit interactions via a 3\textit{d}-5\textit{d} interface has hitherto not been explored. 

The interaction of transition metal oxides with SrIrO$_3$ exhibited very interesting properties, for example, tuning magnetic anisotropy in La$_{1-x}$Sr$_x$MnO$_3$/SrIrO$_3$ superlattices through octahedral rotation\cite{24,25} and inducing metal-insulator transition in LaNiO$_3$ by charge transfer from SrIrO$_3$ \cite{26}. Novel magnetic phases such as spin-glass and skyrmions in SrRuO$_3$/SrIrO$_3$ superlattices were also reported \cite{27}. The interfacial charge transfer driven phenomena like the emergence of magnetism in SrIrO$_3$/SrMnO$_3$ superlattices \cite{1,28} and interfacial re-entrant spin/super spin-glass state has been reported recently in LaMnO$_3$/SrIrO$_3$ bilayer \cite{29}.

In this work, we demonstrate the influence of LaMnO$_3$ layer on magneto-transport and spin-orbit coupling properties at the SrIrO$_3$ interface where LaMnO$_3$ growth condition plays a major role. Our magneto-transport measurements show a tunable and enhanced Rashba spin-orbit coupling at the interface with varying magnetic behaviour of LaMnO$_3$. In addition, X-ray photoelectron spectroscopy (XPS) measurements indicates different fraction of Mn$^{3+}$ and Mn$^{4+}$ valence states in LaMnO$_3$ grown at different oxygen partial pressures, this affects the spin-orbit coupling related parameters at LaMnO$_3$/SrIrO$_3$ interface. Also, interfacial charge transfer from Ir$^{4+}$ to Mn$^{3+}$ and Mn$^{4+}$ from growth variation has not been reported at a 3\textit{d}-5\textit{d} interface, where as individual (Mn$^{3+}$) LaMnO$_3$/SrIrO$_3$ \cite{29} and (Mn$^{4+}$) SrMnO$_3$/SrIrO$_3$ interfaces charge transfer have been reported earlier \cite{1,28}.

Spin Hall magnetoresistance (SMR) has become a versatile tool to probe the nature of magnetic interfaces \cite{30,31}. In our case, LaMnO$_3$ is a magnetic layer and SrIrO$_3$ is a metallic oxide with large spin-orbit coupling \cite{31}. Although SrIrO$_3$ is the best choice for spin-transport studies due to low charge conductivity and large spin-orbit coupling, it has not been thoroughly explored through electrical transport measurements \cite{60}. The angle-dependent magnetoresistance (ADMR) measurements showed signatures of magnetic proximity effect (MPE) in SrIrO$_3$, which is reflected in ADMR magnitude as well. Our study provides a new platform for tuning interfacial effects in TMO heterostructures by interface modifications which may have an impact on designing spintronic devices with an emerging 5\textit{d} quantum material.
\begin{figure*}
\includegraphics[width=16cm,height=16cm,keepaspectratio]{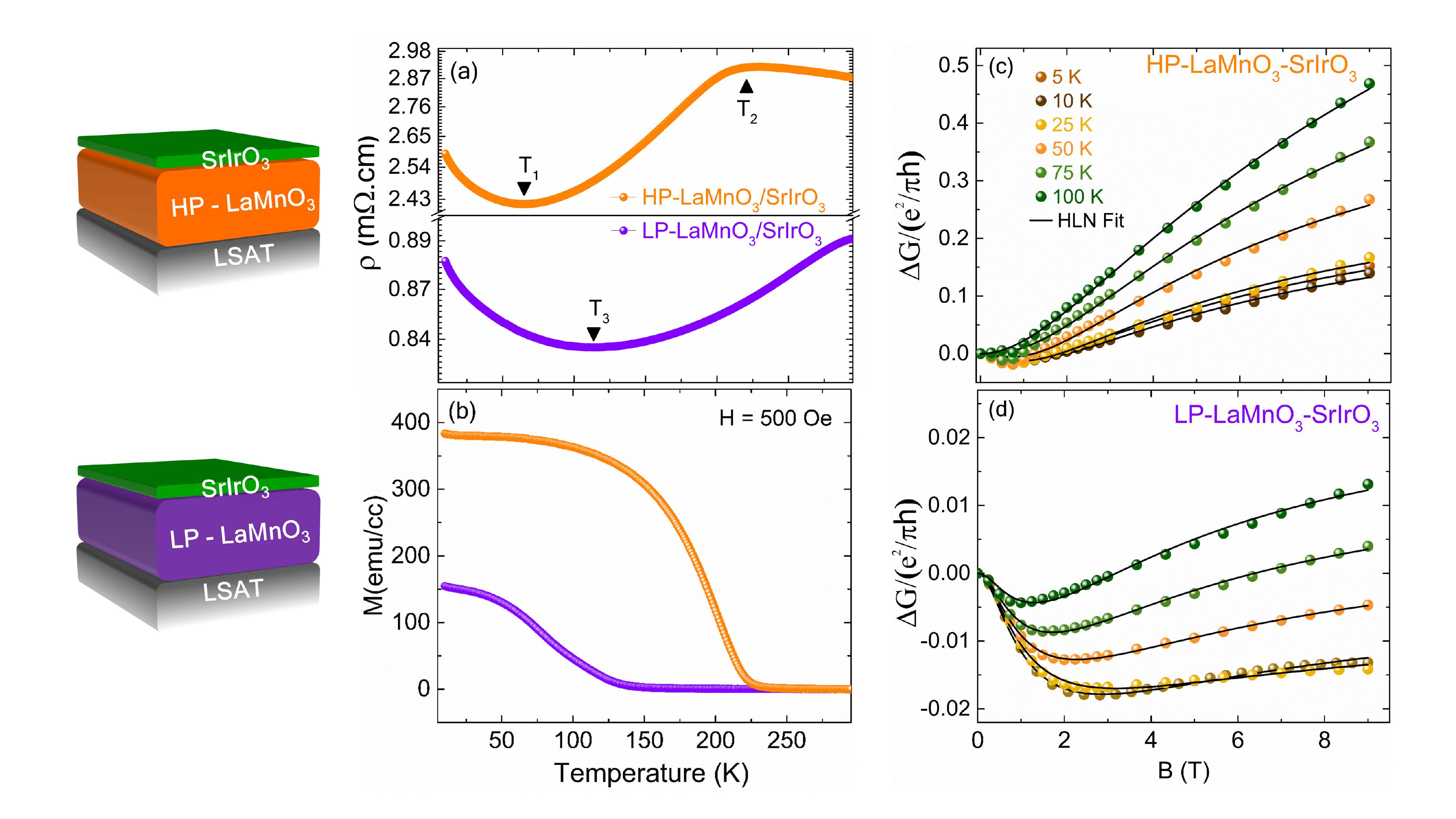}
\caption{\label{fig1} Schematic image of the layered structure of LaMnO$_3$ deposited at  deposited at 37.5 mTorr (HP-LMO-SIO) and 37.5$\times$10$^{-3}$ mTorr (LP-LMO-SIO) bilayer samples (left). (a) Temperature dependence of resistivity ($\rho$) and (b) magnetization (M) at an applied magnetic field of 500 Oe in Field cooled (FC) protocol are demonstrated in (a) and (b) respectively for HP- and LP- LMO-SIO bilayer. (c) and (d) experimental magnetoconductance ($\Delta G$) data (closed colored symbols) as a function of magnetic field (B) $\perp $ interface; measured for different temperatures fitted (solid black curve) by the Hikami-Larkin-Nagaoka equation for HP- and LP- LMO-SIO samples respectively.}
\end{figure*}

\begin{figure*}
\includegraphics[width=14cm,height=14cm,keepaspectratio]{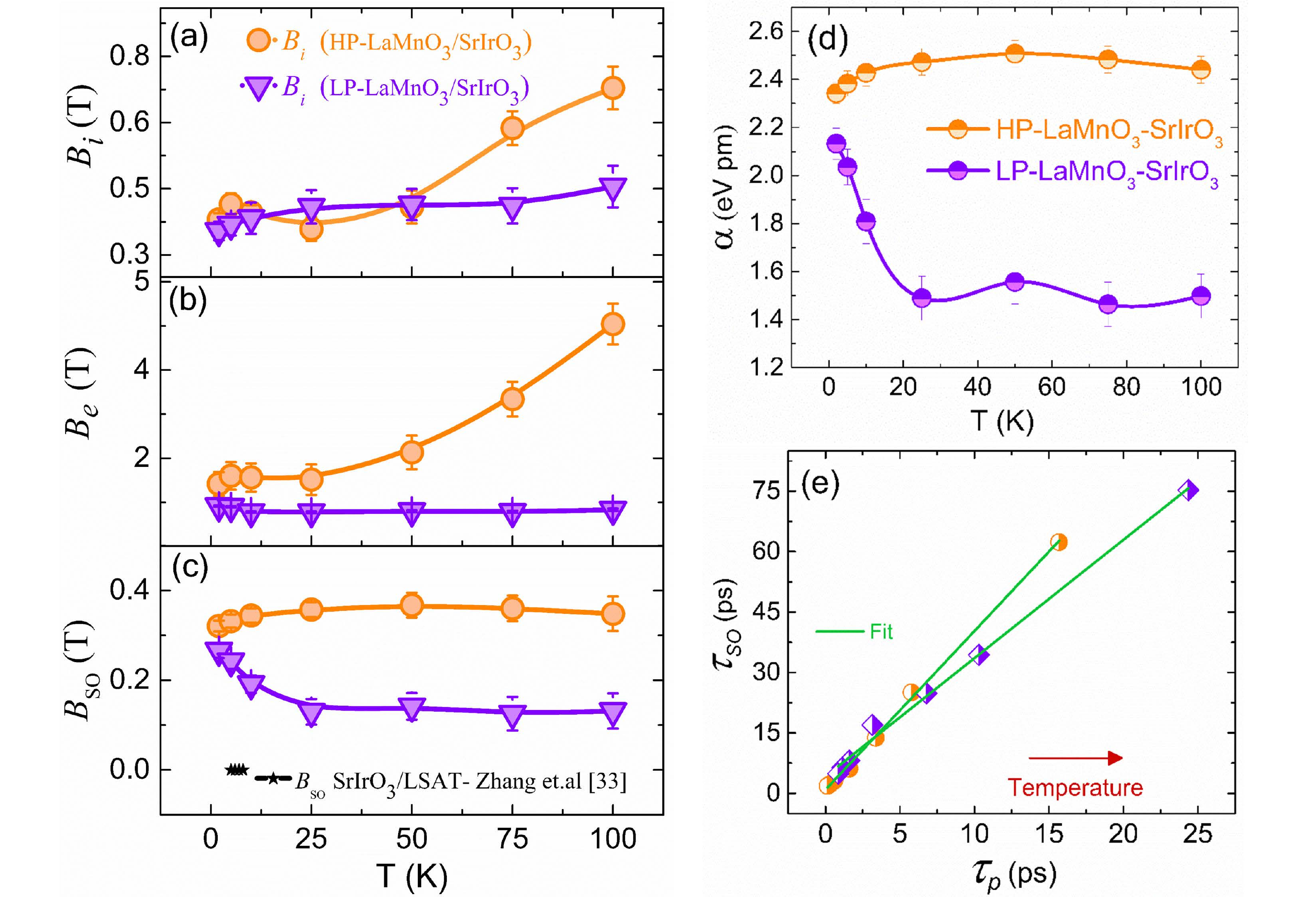}
\caption{\label{fig2} Evolution of fitting parameters  $B_i$ (panel a), $B_e$ (panel b) and $B_{SO}$ (panel c) as a function of temperature for HP- and LP- LMO-SIO sample, also SrIrO$_3$/LSAT (001)\cite{37} has also been plotted for comparison in (c). (d) Rashba spin-orbit coupling ($\alpha$) extracted from $B_{SO}$ values using Eqn.3 is shown as a function of temperature for HP and LP samples. (e) Spin relaxation timescale ($\tau_{so}$) vs momentum scattering timescale ($\tau_{p}$) dependency is consistent with the Elliot–Yafet mechanism for HP- and LP- LMO-SIO samples. Linear fit (solid green colour) where $\tau_{so}$ $\propto$ $\tau_{p}$.}
\end{figure*}
\section{\label{sec:level2}Results and Discussion}

LaMnO$_3$ thin films were grown on (001) oriented (LaAlO$_3$)$_{0.3}$(Sr$_2$TaAlO$_6$)$_{0.7}$ (LSAT) substrates by pulsed laser deposition (PLD) in order to rule out the possibility of polar catastrophe and minimize lattice mismatch. LP-LMO denotes 10 nm thick LaMnO$_3$ grown at low $pO_2$ (37.5$\times$10$^{-3}$ mTorr), similarly HP-LMO denotes 10 nm thick LaMnO$_3$ grown at high $pO_2$ (37.5 mTorr). The LaMnO$_3$/SrIrO$_3$ bilayers with 5 nm thick SrIrO$_3$ deposited on top of HP- and LP- LMO under similar conditions (deposition $pO_2$ = 100 mTorr) are labelled as LP-/HP- LMO-SIO respectively, the detailed growth-related procedure is given in supplementary information Figure S1. The quality of PLD grown LP/HP-LMO samples are confirmed by Atomic Force Microscopy (AFM) and X-ray Diffraction (XRD) studies (Figure S2 and S3). 

The electrical and magnetotransport properties of the samples are investigated using PPMS transport measurement system (more details are given in Supplementary Information). The temperature-dependent resistivity $\rho(T) $ is shown in the Fig.~\ref{fig1}a. To rule out the possibility of electrical conduction channels through LP-/HP- LMO, we have measured resistivity of individual layers grown on LSAT substrates which were highly resistive compared to the SrIrO$_3$ semimetallic layer, as shown in the supplementary Figure S4. The temperature-dependent behaviour of magnetization (field cooling; FC) measured at 500 Oe of these bilayer samples are also shown in Fig.~\ref{fig1}b. The first change of slope in the $\rho $ vs $T$ plot of the HP-LMO-SIO sample was found to be around 50 K (defined as $T_1$, in  Fig.~\ref{fig1}a), which corresponds to weak anti-localization to weak localization crossover usually found in SrIrO$_3$ thin films \cite{36}. The second change of slope near 210 K (defined as $T_2$), which corresponds to the ferromagnetic to paramagnetic transition of LaMnO$_3$, as shown in Figure 1b. Similarly, the change of slope near 120 K (defined as $T_3$) in electrical transport measurements (see Fig.~\ref{fig1}a) also corresponds to Curie temperature ($T_C$) of  the LP-LMO layer (see Fig.~\ref{fig1}b) in the magnetic measurements. A clear shift in the $T_C$  as well as the magnitude of the magnetic moment in both the samples are consistent with earlier reports, ascertaining that the disproportion of Mn$^{3+}$ and Mn$^{4+}$ drives LaMnO$_3$ layer to different magnetic order \cite{22,23}. The oxygen gas atmosphere during the deposition allows oxygen to be absorbed into the lattice, thereby enhancing the formation of Mn$^{4+}$ ions promoting double exchange mediated ferromagnetic ordering in epitaxial HP-LMO thin films \cite{20,22,23}. When the $pO_2$ during LaMnO$_3$ deposition decreases, this enhances formation of increased Mn$^{3+}$ which has smaller magnetic moment compared to Mn$^{4+}$ the LaMnO$_3$ layer evolves to an antiferromagnetic ground state. To quantify Mn$^{3+}$ and Mn$^{4+}$, we have performed X-ray photoelectron spectroscopy (XPS) measurements on HP- and LP- LMO samples (see supplementary Figure S5), which is a surface-sensitive technique. XPS results are in agreement with earlier reports of increased Mn$^{4+}$ content in HP-LMO samples and lower content of Mn$^{4+}$ in LP-LMO samples. Also, to verify the oxygen non-stoichiometry and La/Mn ratio, we have carried out Rutherford backscattering spectroscopy (RBS) in LaMnO$_3$ thin films grown at different $pO_2$. The RBS data (Figure S6) shows that the HP-/LP-LMO samples are nearly stoichiometric, see supplemental material Figure S6(b). The oxygen content in the LaMnO$_3$ thin films are roughly estimated, which has been found to increase for the HP-LMO sample compared to that of the LP-LMO, as shown in supplemental material Figure S6(c).

To further understand the electronic transport behaviour and electron spin relaxation mechanism, magnetotransport measurements were carried out on the bilayer samples at different temperatures as shown in Fig.~\ref{fig1}c and d . The Magnetoconductance (MC) showed a negative to positive crossover which was dominant at low temperatures and this crossover trend becomes weaker with increase in temperature and vanishes near 100 K and 125 K for LP-/HP- LMO-SIO respectively. At low magnetic fields the negative MC component is dominant, and at high magnetic fields the positive MC component is dominant. Negative MC (at temperatures below 10 K) reported in ultrathin SrIrO$_3$ films grown on compressively strained LSAT and STO substrates arises due to the competition between weak localization (WL) and strong spin-orbit coupling based weak anti-localization effects (WAL) \cite{37}. Usually, from various reports, the crossover from negative to positive MC arises in ultrathin SrIrO$_3$ thin films in the temperature range of 7-10 K \cite{13,36,37}. However, we observed a crossover in MC at low magnetic fields in the temperature range of 100 to 125 K for both HP- and LP- LMO-SIO, as shown in Fig.~\ref{fig1}c and d. SrIrO$_3$ grown on HP- and LP- LMO show different temperature dependence in the crossover of MC from positive MC to negative MC. In addition, the shape of MC has also been found to change for both samples. To investigate this scenario in terms of spin-orbit coupling in the LP- and HP- LMO-SIO layer, we used Hikami-Larkin-Nagaoka (HLN) equation \cite{38} to fit the MC data.
\newline
\newline
\begin{widetext}
\begin{equation}
    \frac{\Delta \sigma(B)}{G_0}= -\Psi \left({\frac{1}{2}+\frac{B_e}{B}}\right)+ \frac{3}{2} \Psi \left({\frac{1}{2}+\frac{B_i +B_{so}}{B}}\right)+-\frac{1}{2} \Psi \left({\frac{1}{2}+\frac{B_i}{B}}\right)- \ln \left({\frac{B_i +B_{so}}{B_e}}\right) - \frac{1}{2} \ln \left({\frac{B_i +B_{so}}{B}}\right)
    \label{eq1}
\end{equation}
\end{widetext}
In the Eq. (1) $\psi $ is digamma function and G$_0$ is the universal conductance constant; 1.2 x 10-5 S. $B_e$, $B_i$ and $B_{so}$ represents effective fields of elastic, inelastic, and spin-orbit coupling induced scattering terms, respectively. HLN equation best describes the competition between spin-orbit coupling and weak localization. MC behaviour in LP/HP-LMO-SIO is in good agreement with the HLN model in the temperature range between 5 K and 100 K. We could extract different scattering parameters as a function of temperature, as shown in Fig.~\ref{fig2}a-c. The magnitude of these parameters obtained for our samples is one order of magnitude higher in comparison to the direct SrIrO$_3$ layer grown on LSAT with similar deposition conditions.

The elastic scattering field (B$_e$) which is one order of magnitude higher compared to B$_i$ and $B_{so}$ fields which are in agreement with the fact that the electronic transport is dominated by two-dimensional (2D) weak localization \cite{39}. We could see that the parameter $B_i$ has comparable magnitudes in both LP-/HP- LMO-SIO samples and follows a similar temperature dependent trend. However, HP-LMO-SIO has higher B$_e$ values compared to the LP-LMO-SIO, which agrees with the interaction of conduction electrons in SrIrO$_3$ with the magnetic moment of interfacial Mn spins. In HP-LMO-SIO the magnetic moment is higher compared to LP-LMO-SIO, hence the higher magnitude of B$_e$ in the HP-LMO-SIO sample can be attributed to electrons screened due to interfacial Mn spins.

In the case of $B_{so}$, both samples exhibit different behaviour as function of temperature, and the LP-LMO-SIO show a decrease in $B_{so}$ and saturates above 25 K. Whereas the $B_{so}$ parameter increases till 25 K and saturates afterwards for the HP-LMO-SIO. The magnitude of $B_{so}$ is one order higher compared to SrIrO$_3$ directly grown on LSAT. The $B_{so}$ parameter is directly related to the induced spin-orbit coupling at the SrIrO$_3$ layer. Therefore, the temperature-dependent trend of $B_{so}$ and $B_{e}$ parameters point to the fact that the Mn spins at the interface and their magnetic moment plays a vital role in tuning the spin-orbit coupling at the interface.

The role of Mn spins on the scattering of electrons at the interface could be arising from an internal electric field generated due to charge transfer from Ir ions to Mn ions. Recent report of Huang \textit{et al.} on LaMnO$_3$/SrIrO$_3$ superlattices showed the internal electric field arising from the strain induced in the Ir-O-Ir bond angle, which is having a Rashba-like character, at the LaMnO$_3$-SrIrO$_3$ interface \cite{40}. Rashba interactions caused by broken mirror symmetry, and in particular by the associated electric field perpendicular to the SrIrO$_3$ interface induces orbital and lattice polarization due to asymmetric interfacial structure of LaMnO$_3$-SrIrO$_3$ interface \cite{40}. The temperature-dependent Rashba spin-orbit coupling has earlier studied in several semiconductor heterostructures \cite{41}. The higher-order terms in Rashba spin-orbit Hamiltonian is found to be origin of this temperature dependence \cite{42}. In case of SrIrO$_3$, temperature dependence of Rashba spin-orbit coupling  is reported to arise from changes in Landé g factor which is affected by temperature \cite{37}. To investigate the role of Rashba spin-orbit coupling in magnetotransport at the LaMnO$_3$-SrIrO$_3$ interface, the Rashba coefficients for LP-/HP- LMO-SIO were obtained as a function of temperature from $B_{so}$ parameter. The $B_{so}$  parameter is related to the Rashba spin-orbit coupling coefficient as,\cite{37,43,59}

\begin{equation}
    \alpha= \frac{(e\hbar^3 B_{so})^{\frac{1}{2}}}{m_*}
    \label{eq2}
\end{equation}

\begin{figure*}
\includegraphics[width=16cm,height=16cm,keepaspectratio]{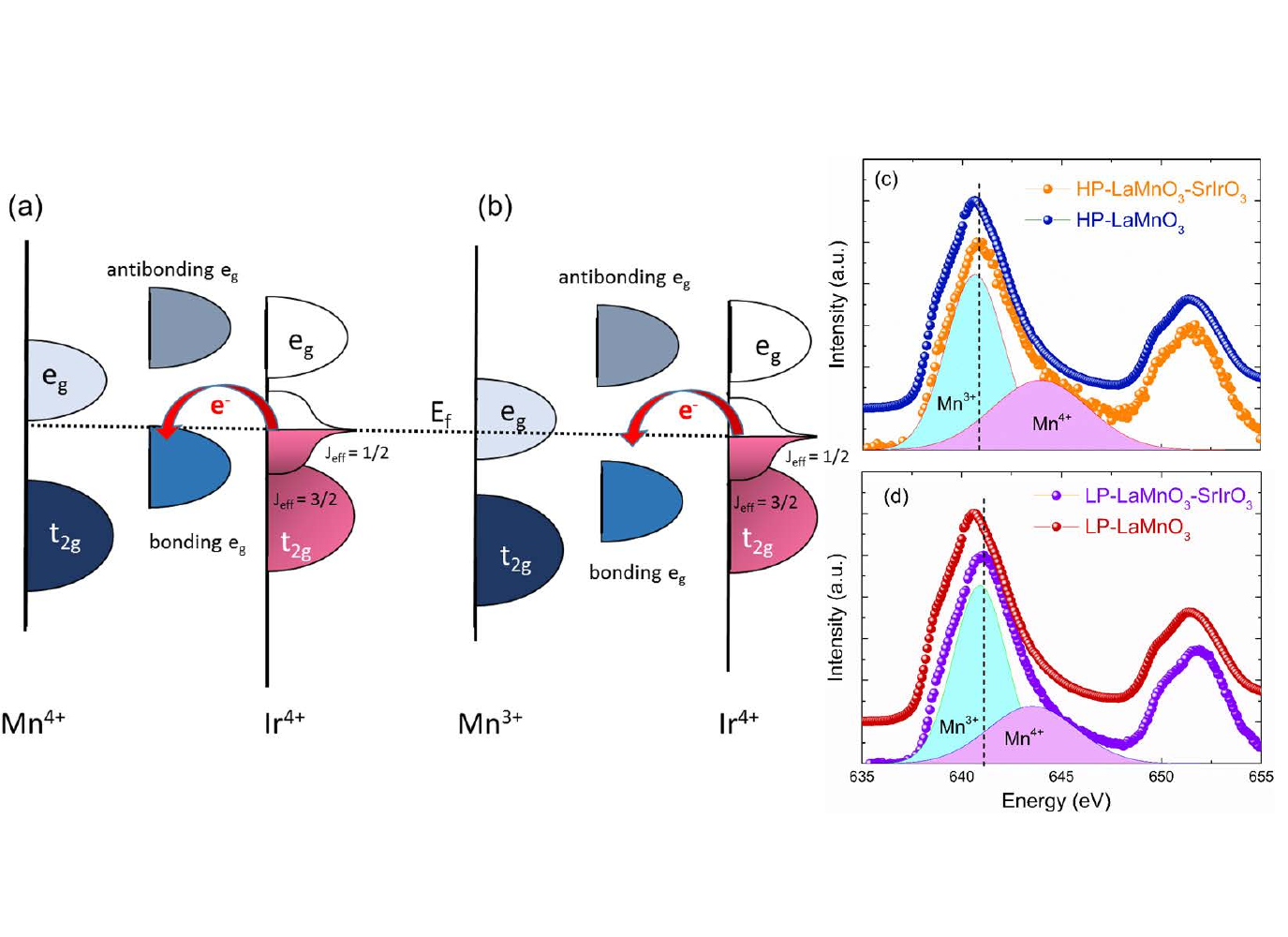}
\caption{\label{fig3}Schematic representation of charge transfer mechanism for Ir$^{4+}$ to Mn$^{4+}$ (panel a) and Ir$^{4+}$ to Mn$^{3+}$ (panel b). (c) and (d) XAS spectra around Mn $L_{2,3}$ edge of HP- and LP- LMO-SIO samples along with corresponding HP- and LP- LMO samples without SrIrO$_3$ layer. The respective Mn valence state position, as deconvoluted Mn \textit{L-3} edge is shown for Mn$^{3+}$ and Mn$^{4+}$ in the bilayer as a shaded area (valence states are quantified by XPS spectra analysis (see supplementary Fig. S4), the shaded area is for representation). The dashed black lines in the figure are guide to the eye based on the shift of the Mn-$L$ edge position of LP- and HP- LMO-SIO with their corresponding LaMnO$_3$ samples.}
\end{figure*}

Here $m^*$ is the effective mass, in case of SrIrO$_3$ $m^*$ $\sim$7$m_o$ ($m_o$: the mass of an electron) \cite{37,43} $e$ is the elementary charge and $\hbar $ is the reduced Planck’s constant. The Rashba spin-orbit coupling coefficients ($\alpha$ ; eVpm) are plotted as a function of temperature for both samples, as shown in Fig.~\ref{fig2}d. The value of $\alpha$ has been obtained for SrIrO$_3$ thin films grown on compressively strained LSAT (001) and STO (001) substrate by Zhang \textit{et al.} \cite{37}, which exactly matches with the single SrIrO$_3$ layer grown on LSAT substrates. In both LP-/HP- LMO-SIO samples, the Rashba spin-orbit coupling coefficient is found to be even higher, with a 10\% increase compared to SrIrO$_3$ directly grown on LSAT substrates. This enhanced Rashba spin-orbit coupling  is due to charge transfer, depending on the valence state of underlying Mn, which shows different temperature dependence at low temperatures. Moreover, HP-LMO-SIO shows a weak temperature dependence compared to LP-LMO-SIO at low temperatures though both saturates at high temperatures (above 25 K). The interfacial coupling between different magnetically ordered LaMnO$_3$ also provides an impact on the coupling between the spin-orbit coupled state of Ir$^{4+}$. As we know, changing the growth pressure on LaMnO$_3$ significantly affects the lattice constant of LaMnO$_3$ layer \cite{20}. LaMnO$_3$ grown under low $pO_2$ were partially relaxed with the lattice constant a$_{LP}$ = 0.400 nm. On the other hand, for thin films grown under oxidizing atmospheres are found to be compressively strained (-0.63\%), with lattice constant a$_{HP}$ = 0.392 nm (shown in supplementary Figure S3). It has been found that SrIrO$_3$ grown on LaMnO$_3$ is strained due to the strain in LaMnO$_3$ lattice. This may lead to change in Ir-O-Ir bond angle (lattice polarization). The IrO$_6$ octahedral rotation due to strain in the Ir-O-Ir bond angle  has also been found to enhance interfacial charge transfer\cite{28}, that may enhance the electric field responsible for Rashba spin-orbit coupling.

To get more insights about the influence of interface induced Rashba spin-orbit coupling in spin relaxation mechanism in these bilayers, we considered two commonly observed mechanisms: the D’yakonov-Perel (DP) and the Elliot-Yafet (EY) mechanisms \cite{44}. The DP type spin relaxation arises in systems that lack inversion symmetry, in which the electron spin precesses in an effective magnetic field with its direction changing after each scattering event \cite{45,46}. Depending on whether it is bulk or interface, the DP mechanism has Dresselhaus and Rashba type contributions respectively \cite{47,48}. On the other hand, the EY mechanism originates from spin-orbit coupling induced spin dephasing due to electron-phonon coupling or interfacial defects \cite{44}. Apart from this in thin films, there could be other contributions to EY mechanism such as scattering events at the grain boundary, oxygen vacancy induced defects and lattice dislocations \cite{49,50}. Both DP and EY mechanisms can be identified by the relation between spin scattering timescale $\tau_{so}$ and momentum scattering timescale $\tau_p$. If the $\tau_{so}$ scales linearly with $\tau_p$ then the dominant mechanism is EY, and if it is inversely proportional, the DP type is the dominant mechanism \cite{44}. The magnitudes of $\tau_{so}$ will be greater than $\tau_p$ in semiconducting systems where the spin orbit coupling is usually weak which results in EY mechanism\cite{44}. This is in contrast with materials with large spin-orbit coupling like Au, where the magnitudes of $\tau_{so}$ and $\tau_p$ are comparable \cite{51}. A recent theoretical report by Kiss \textit{et al.} with a generalized theory of EY mechanism in materials with large spin-orbit coupling argues that if the spin-orbit coupling energy ($\xi$) is comparable with the coulombic correlation energy ($U$), the $\tau_p$ values can approach $\tau_{so}$ values \cite{52}.

In HP- and LP- LMO-SIO samples the spin relaxation timescale $\tau_{so}$ and momentum relaxation ($\tau_p$)  timescale were estimated from $B_{so}$ and $B_{e}$ fields respectively through,\cite{44}

\begin{subequations}
    \begin{align}
      \tau_{so}= \frac{\hbar}{4e B_{so} D} 
      \label{eq3a}\\
      \tau_{p}= \frac{\hbar}{4e B_{e} D}  
       \label{eq3b}
    \end{align}
  \end{subequations}
Here $D$ is the diffusion coefficient, which is calculated from the mobility deduced from Kohler’s equation near small magnetic fields \cite{53,54}. To investigate the correlations between $\tau_{so}$ and $\tau_{p}$ in the HP- and LP- LMO-SIO samples, $\tau_{so}$ and $\tau_{p}$ with increase in temperatures are plotted in Fig.~\ref{fig2}e. Elliott-Yafet (EY) mechanism has recently been reported as the dominant spin relaxation mechanism in SrIrO$_3$ thin films \cite{55}. Here, $\tau_{so}$ is directly proportional to $\tau_{p}$ which shows that the dominant spin relaxation mechanism is the EY mechanism in both HP-LMO-SIO and LP-LMO-SIO samples as shown in Fig.~\ref{fig2}e. As the interface showed enhancement in Rashba spin-orbit coupling, the much-anticipated spin relaxation mechanism was DP, instead we observed a dominant EY mechanism with 100\% enhanced $\tau_{p}$ values compared to previous reports \cite{55}. Our observation of EY mechanism in spin relaxation and relatively higher and comparable magnitude of $\tau_{so}$, in the LP- and HP- LMO-SIO agrees with theoretical report of the Kiss \textit{et al.} on the role of spin-orbit coupling in enhanced and comparable relaxation time scales in the EY mechanism in strongly correlated systems\cite{52}. Though the magnitude of Rashba spin-orbit coupling is higher in the LaMnO$_3$-SrIrO$_3$ interface, contribution from DP mechanism might get suppressed due to the large spin-orbit coupling. The charge transfer induced electric field may induce scattering centers resulting in an EY type spin relaxation mechanism. The fact that the slope of $\tau_{so}$ vs $\tau_{p}$ plot (Fig.~\ref{fig2}e) for HP-LMO-SIO sample is higher compared to the LP-LMO-SIO sample underlines the above argument, this also reaffirms the relatively higher contribution of spin-flip scattering at the interface due to higher magnitude of HP-LMO-SIO magnetic moment. The enhanced magnitudes of $\tau_{so}$ and $\tau_{p}$ is related to the interfacial modifications induced spin-orbit coupling interactions, which needs to be understood further in terms of charge transfer effects. 

\begin{figure*}
\includegraphics[width=16 cm,height=16cm,keepaspectratio]{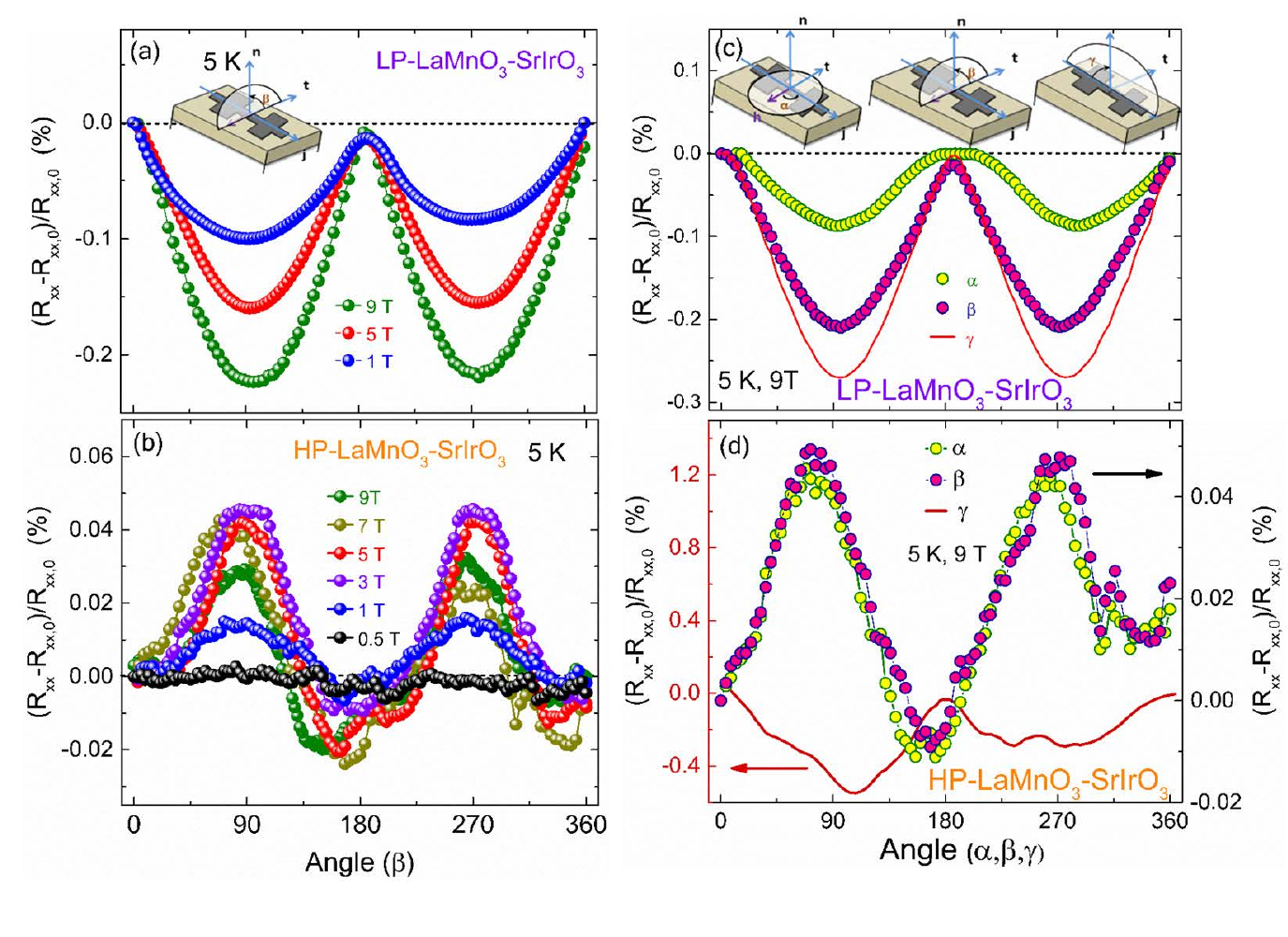}
\caption{\label{fig4}(a) and (b) The angle-dependent magnetoresistance (ADMR) measurements carried out as a function of the magnetic field for LP- and HP- LMO-SIO samples respectively in $\beta$ rotational plane (Schematic of the $\beta$ rotational measurement configuration in inset). Figures (c) and (d) ADMR measurements as a function of different rotational configurations. The rotational planes are shown as an inset in (c). The angles $\alpha,\beta,\gamma$ are defined as the angle subtended between the current direction $j$ with respect to the magnetic field rotation, and $n$ is represented as a  direction cosine normal to the surface. $\alpha$ points to the in-plane (IP) rotation of magnetic field with respect to $n$. $\beta$ represent out of plane (OOP) rotation direction lying in the plane perpendicular to the current direction $j$. $\gamma$ shows the OOP direction with respect to the current direction plane $j$.}
\end{figure*}

In the case of bulk LaMnO$_3$, under the crystal field created by oxygen 2p states in octahedrally coordinated MnO$_6$, the five Mn $3d$ levels split into low-energy $t_{2g}$ triplet and high-energy $e_{g}$  doublet levels, as sketched in Fig.~\ref{fig3}a and b. In general, in hole-doped LaMnO$_3$ systems, the Mn ions are in a mixed trivalent ($3d^4$) and tetravalent ($3d^3$) states \cite{29}. In the case of Mn$^{4+}$ the $e_{g}$ orbitals are empty and singly occupied for Mn$^{3+}$. In our case, we have mixed-valence states with varying amount of Mn$^{3+}$ and Mn$^{4+}$ for HP- and LP- LMO-SIO samples. Further, the $e_{g}$ states of Mn$^{3+}$ and Mn$^{4+}$ couples with interfacial $e_{g}$ states of Ir$^{4+}$ to give rise to molecular orbitals with energetically lower lying bonding and upper lying antibonding levels \cite{29}. This coupling of $e_{g}$(3$z^2$-$r^2$ bonding orbital) states at the interface promotes charge transfer from Ir$^{4+}$ to Mn$^{3+}$ and Mn$^{4+}$ states. To experimentally ascertain the charge transfer at the interface, X-ray absorption spectroscopy (XAS) study in total-electron-yield (TEY) mode has been performed on HP- and LP- LMO-SIO bilayer samples and, HP- and LP-LMO without SrIrO$_3$ layer as illustrated in Fig.~\ref{fig3}c and d. At first, XAS data rules out the existence of any other valence states other than Mn$^{3+}$ and Mn$^{4+}$ in LaMnO$_3$. The LP-LMO-SIO bilayer sample shows a shift of 1.54 eV towards higher energy compared to LP-LMO and similarly HP-LMO-SIO shows a shift of 0.25 eV compared to HP-LMO. The LP-LMO-SIO sample shows a pronounced shift due to a predominant concentration of Mn$^{3+}$ ions over Mn$^{4+}$ ions as seen in XPS spectra (see Figure S5). Since the Mn$^{3+}$ bonding orbital lies much lower to the Fermi level compared to Mn$^{4+}$ (sketched in Figure 3a and 3b) the Mn$^{3+}$ contributes predominantly to the charge transfer process, this observation is in compliance with recently reported LaMnO$_3$/SrIrO$_3$ superlattices \cite{40}. Also the charge transfer has two competing contributions arising from strain in IrO$_6$ octahedra and due to overlap of low lying $e_{g}$ bonding orbitals in Mn$^{3+}$ compared to Mn$^{4+}$ at the interface. Our data shows the charge transfer being responsible for EY type spin relaxation mechanism in LMO-SIO interface. However, it is still necessary to use other complimentary experimental methods to understand further the variation in Landé g-factor of SrIrO$_3$ in LP-/HP- LMO-SIO heterostructures as a function of temperature.

We have further carried out angle-dependent magnetoresistance (ADMR) measurements to understand the charge transfer effects on transport behaviour in SrIrO$_3$ due to the interfacial Mn spins in HP- and LP-LMO-SIO samples. The ADMR(\%)$=$[R($\alpha$,$\beta$,$\gamma$)$-$ $R(0))/R(0)$] investigated as a function of magnetic field and rotational planes at 5 K, where $R(0)$ is the resistance when magnetic field is normal to sample surface and $R(\alpha,\beta,\gamma)$ is the resistance with respect to each rotational plane $(\alpha,\beta,\gamma)$ as shown in the Figure 4c inset. We observed ADMR signals with a phase shift for LP-LMO-SIO compared to HP-LMO-SIO sample. In particular, we observe a distinct magnetoresistance trend in LP-LMO-SIO sample as [MR($\gamma$) $>$ MR($\beta$)$>>$MR($\alpha$)](Fig.4c) compared to spin Hall magnetoresistance (SMR) which has the form [MR($\alpha$)=MR($\beta$)$>>$ MR($\gamma$)]. This ADMR trend of LP-LMO-SIO does not comply with the conditions meant for anisotropic magnetoresistance (AMR), $i.e.$ [MR($\alpha$)= MR ($\gamma$)$>>$ MR($\beta$)]. This is also distinct from the recently reported proximity induced magnetoresistance (PMR) for which the condition is [MR($\beta$)$=$ MR($\gamma$)$>>$MR($\alpha$)] \cite{56}. On the other hand, in the case of the HP-LMO-SIO the trend points to [MR($\gamma$)$>$ MR($\beta$)= MR($\alpha$)] (Fig.4d), this does not resemble to any above mentioned magnetoresistance models. Additionally, the observed ADMR data is quite different from reported MPE (magnetic proximity effect) in SMR of ferromagnets (FM)\cite{57} and MPE in SMR of antiferromagnets (AFM) \cite{58}.

LaMnO$_3$ orients as an A-type antiferromagnet in bulk and thin films, as shown in earlier reports using scanning SQUID microscopy \cite{19}. In the case of fully relaxed LaMnO$_3$ thin films, the intra-plane exchange interaction is ferromagnetic, whereas the inter-plane exchange interaction is antiferromagnetic, which may lead to an A-type antiferromagnetic ordering. However, in our case we have strained epitaxial films, that have ferromagnetic exchange interactions that results in overall ferromagnetic ordering. In LP-LMO-SIO samples, we expect the overall magnetic ordering is antiferromagnetic and the trend and magnitude of ADMR correspond to the effect of induced magnetism at the interface, such that magnetization rotation is reflected in the ADMR data. This observation were in agreement with the field-dependent changes in ADMR signals of LP-/HP- LMO-SIO samples. In the case of LaMnO$_3$-SrIrO$_3$ bilayers, there is a mixed effect from SMR as well as proximity-induced magnetism (MPE) arising from charge transfer at the interface. The field-dependent ADMR (Fig.~\ref{fig4}a and b) in HP-LMO-SIO sample show an amplified signal at 3 T, and it decreases with increasing magnetic field strength, which suggests the existence of competing domains which percolates with increasing magnetic field in LaMnO$_3$.  In the LP-LMO-SIO sample case, we have observed a non-saturating ADMR which is due to the presence of antiferromagnetic order that prevails with the magnetic field and MPE being a dominant contribution arising from induced magnetism in LaMnO$_3$-SrIrO$_3$ interface. Further experimental and theoretical studies are required to understand the domain structure of LaMnO$_3$ with MPE effects, to extract the spin-charge conversion at LaMnO$_3$-SrIrO$_3$ interface using a theoretical macrospin model to explore such a complicated magnetic structure.

\section{\label{sec:level3}Conclusion\protect}
In summary, we have explored the 3\textit{d}-5\textit{d} interface interactions through tunability of the magnetic order in LaMnO$_3$ by varying the oxygen partial pressure during LaMnO$_3$ deposition. The X-ray spectroscopy measurements indicate the changes in the magnetic ordering of LaMnO$_3$ can be attributed to the creation of multiple valence states. A tunable and enhanced Rashba spin-orbit coupling is estimated at LP- and HP-LMO-SIO as a function of temperature from magnetoconductance measurements that arises from electric field generated due to strain in Ir-O-Ir bond angle as well as interfacial charge transfer from Ir$^{4+}$ to Mn$^{3+}$ and Mn$^{4+}$. The spin relaxation mechanism at LaMnO$_3$-SrIrO$_3$ interface is observed to follow the Elliott-Yafet mechanism. The spin relaxation parameters are in agreement with the generalized theory of EY mechanism for materials with large spin-orbit coupling. The contribution of Mn spins in LaMnO$_3$ on electronic transport was further probed using ADMR measurements, which reflects the magnetic order of underlying LaMnO$_3$ and charge transfer induced magnetism at LaMnO$_3$-SrIrO$_3$ interface. The evolution of these phenomena is attributed to the 3\textit{d}-5\textit{d} interface electronic correlation and the Rashba spin-orbit coupling at the LaMnO$_3$-SrIrO$_3$ interface. In conclusion, the present results provide a novel platform of 3\textit{d}-5\textit{d} oxide interface engineering and raises possibilities in tuning these interface interactions to optimize spin transport in emerging quantum material SrIrO$_3$.

\begin{acknowledgments} 
	M.S.R. and K.S would like to acknowledge the SERB–DST Extra Mural Funding for the purchase of a horizontal rotating holder (Quantum Design) from the project EMR/2017/002328. The authors would like to acknowledge Dr. Matthias Opel and Dr. Matthias Althammer of Walther Meissner Institute Garching, Technical University of Munich (TUM) Germany for insightful discussions and suggestions in the modification of the manuscript. The authors would like to acknowledge Professor Mahendran and his student Amit Chanda of National University of Singapore (NUS) for their help in measurements. This work was financially supported by the SERB-DST (EMR/2017/002328).  The research at NUS is supported by the Agency for Science, Technology and Research (A*STAR) under its Advanced Manufacturing and Engineering (AME) Individual Research Grant (IRG)  (A1983c0034), the National University of Singapore (NUS) Academic Research Fund (AcRF Tier 1 Grant No. R-144-000-391-144 and R-144-000-403-114)  and the Singapore National Research Foundation (NRF) under the Competitive Research Programs (CRP Award No. NRF-CRP15-2015-01). H.J. would like to thank NUS Graduate School of Integrative Science and Engineering (NGS) for fellowship. 
\end{acknowledgments}

\nocite{*}
\bibliographystyle{apsrev4-2}
\bibliography{apssamp} 

\end{document}